\documentclass[twoside]{article}
\usepackage{fleqn}
\usepackage{rrh}
\usepackage{epsfig}
\usepackage{graphics}
\def\benum{\begin{enumerate}}
\def\eenum{\end{enumerate}}

\def\ben{\begin{equation}}
\def\ba{\begin{array}}
\def\bea{\begin{eqnarray}}
\def\bec{\begin{center}}

\def\een{\end{equation}}
\def\eea{\end{eqnarray}}
\def\enc{\end{center}}
\def\ea{\end{array}}
\def\btab{\begin{table}}
\def\btabu{\begin{tabular}}
\def\etab{\end{table}}
\def\etabu{\end{tabular}}
\def\bit{\begin{itemize}}
\def\eit{\end{itemize}}
\def\bef{\begin{figure}[htb]}
\def\befh{\begin{figure}[!h!]}
\def\enf{\end{figure}}

\def\la{\langle}
\def\ra{\rangle}

\def\bsig{\mbox{\boldmath $\sigma$}}

\def\bD{\mbox{\boldmath $\D$}}

\def\gb{\beta}
\def\D{\Delta}
\def\de{\delta}

\def\b1{{\bf 1}}

\def\bx{{\mbox{\boldmath $x$}}}
\def\by{{\mbox{\boldmath $y$}}}

\def\bp{{\mbox{\boldmath $p$}}}

\def\cos{\hbox{cos}\:}

\def\nn{\nonumber}
\def\bb{\left(}
\def\eb{\right)}

\def\bD{{\bf D}}

\def\rb{\raisebox{3mm}[0pt]}

\usepackage{epsfig,latexsym}
\def\bem{\begin{minipage}}
\def\enm{\end{minipage}}

\title{
NRQCD on an anisotropic lattice
\thanks{Presented by R.R. Horgan. This work is supported in part 
by the Leverhulme Trust grant no. F618C and by PPARC Research Grant GR/L56039.}
}
\author{
I.T. Drummond R.R. Horgan T. Manke H.P. Shanahan
\address{DAMTP, University of Cambridge, Silver Street, 
Cambridge, England CB4 4SL}
}
\begin{document}
\begin{abstract}
We present preliminary results for the $\Upsilon$ spectrum on an anisotropic 
lattice using the improved $O(mv^6)$ NRQCD Hamiltonian. We find that accurate results 
can be obtained in moderate computer times and that they agree with earlier
results on an isotropic lattice.
\end{abstract}

\maketitle
\section{\label{ani}\bf The anisotropic lattice}

\enlargethispage{-7truemm}
On an anisotropic lattice an improved action is used in which the
coupling in the ``time'' direction is tuned so that the temporal
lattice spacing $a_t$ is much smaller than the spatial lattice spacing
$a_s$, thus allowing refined measurements of mass differences
whilst allowing a coarse spatial lattice. The bare anisotropy $\chi$ is a parameter 
in the action which determines the renormalized anisotropy $\chi_R = a_s/a_t$. 
We use the improved action given by  \cite{morn0,alea}:

\bea 
\lefteqn{{S}_{n}=-\gb \sum_{x,\,s>s^\prime} \chi^{-1} \left\{  
\frac{5}{3}  \frac{P_{s,s^\prime}}{u_s^4} - \frac{1}{12} \frac{R_{ss,s^\prime}}{u_s^6}\right. -}\nn\\
\lefteqn{\left.\frac{1}{12} \frac{R_{s^\prime s^\prime,s}}{u_s^6}\right\} - 
\gb \sum_{x,\,s} \chi\,\left\{ \frac{4}{3}  
\frac{P_{s,t}}{u_s^2 u_t^2} - \frac{1}{12} \frac{R_{ss,t}}{u_s^4 u_t^2} \right\}~.}\nn
\eea
Here $s,s^\prime$ run over spatial directions, $P_{s,s^\prime}$ is a $1\times 1$ plaquette,
$R_{s^\prime s^\prime,s}$ is a $2\times 1$ plaquette, and $u_s$ and $u_t$ are tadpole improvement
parameters which are determined self-consistently. The value of $\chi_R$ is measured, for example, 
by computing the static quark potential in both the coarse and fine directions and we suggest below 
an alternative method to measure $\chi_R$.  We use lattices whose parameters were given to us by Alford et al.: 
$\gb = 1.8,~\chi = 4,~\chi_R = 3.815(10)~,u_s = 0.7255~,u_t = 0.9812,~$ where the tadpole 
coefficients are given by the respective mean link values in the Landau gauge. Typical lattice sizes used were 
$8^3\times 40$ and $6^3 \times 60$.

Whilst we have worked with the values given above for $u_s$ and $u_t$ we have also
implemented a fourier accelerated self-consistent calculation of the Landau gauge
definition of these parameters. The definition used for Landau gauge is
to maximize
\[
\sum_{x\;\mu}\,{1 \over u_\mu\,a_\mu^2}\mbox{ReTr}\left\{ U_\mu(x)-
{1 \over 16u_\mu}U_\mu(x)U_\mu(x+\hat{\mu})\right\}.
\]
with respect to gauge transformations, and define $u_\mu = \la U_\mu \ra_{Landau}$.
this procedure is very time-consuming. For example, on $8^3\times 32$ to fix the gauge 
to 1 in $2\cdot 10^{-4}$ requires between 50 and 200 iterations taking, on average, 
30 minutes on a 533 MHz DEC PC. To obtain better than 1\% accuracy requires about 50 
such measurements. We have made preliminary determinations of $u_s$ and $u_t$ on 
$8^3\times 32$ and $4^3\times 16$ and find a lattice size dependence:
\bea
\lefteqn{4^3\times 16:~~u_s~=~0.680(2)~,~u_t~=~0.9792(3),}\nn\\
\lefteqn{8^3\times 32:~~u_s~=~0.707(3)~,~u_t~=~0.9811(2).}\nn
\eea
The $8^3$ values are close to those of Alford et al. and the discrepancy should
have little effect on our results.
 
\section{\bf NRQCD to $O(mv^6)$}

\enlargethispage{-7truemm}
In NRQCD the inversion problem of the fermion matrix is an initial value problem.
We use the evolution equation along the Euclidean time-direction defined by:
\bea
\lefteqn{G(\bx,t+1;\by)=}\nn\\
\lefteqn{~~~~{\cal U}(n,t)\;(1-a_t\de H)\;G(\bx,t;\by)~,~~~~~ t>0} \nn\\
\lefteqn{G(\bx,1;\by)={\cal U}(n,0)\;S(\bx,\by)~,} \nn\\
\lefteqn{{\cal U}(t,n)=\bb 1-{a_tH_0 \over 2n} \eb^n U^\dagger_t \bb 1-{a_tH_0 \over 2n} \eb^n~.} \nn\\
\eea
Here $S(\bx,\by)$ is the source at the first timeslice, $t=0$~. We have 
$S(\bx,\by) = \de^{(3)}(\bx,\by)$ for a single quark at the origin but we also
evolve a quark with a smeared source centred at $\by$. For smearing on a length scale
$l$ the extended source is defined by
\ben
S(\bx,\by;l)=\bb1+{l^2\bD^2 \over 4m}\eb^m\,\de^{(3)}(\bx,\by),\label{smear_S}
\een
where $\bD$ is the covariant derivative on $\bx$. Typically, $l=1,~m=10$~.
To construct $P$-wave meson states we also need single quark propagators evolved
from a source of the form 
\ben
P_i(\bx,\by;l)=\bb1+{l^2\bD^2 \over 4m}\eb^m\,D_i\de^{(3)}(\bx,\by).\label{smear_P}
\een

The NRQCD Hamiltonian to $O(mv^6)$ is given by \cite{leea,morn}

\bea
\lefteqn{H_0~=~-~\frac{\Delta^2}{2m_b} ~,} \nn \\
\lefteqn{\de H ~=~ -~c_0 \frac{\Delta^4}{8m_b^3}-c_1 \frac{1}{2m_b} \sigma\cdot g{\bf B}}\nn\\
&+&c_2 \frac{i}{8m_b^2}(\Delta \cdot g{\bf E} - g{\bf E} \cdot \Delta) \nn\\
&-&c_3 \frac{1}{8m_b^2} \sigma \cdot ( \tilde{\Delta} \times g{\bf E} - g{\bf E} \times \tilde{\Delta})\nn\\
&-&c_4 \frac{1}{8m_b^3}\{\Delta^2,\sigma \cdot g{\bf B}\} \nn \\  
&-&c_5 \frac{3}{64m_b^3}\{\Delta^2, \sigma \cdot ( \Delta \times g{\bf E} - g{\bf E} \times \Delta)\}\nn\\
&-&c_6 \frac{i}{8m_b^3} \sigma \cdot g{\bf E} \times g{\bf E} \nn\\
&-&c_7 \frac{a_s\Delta^4}{16n\chi_R m_b^2}~+~c_8 \frac{a_s^2 \Delta^{(4)}}{24m_b} ~. \label{H}
\eea

The fields $\bf E$ and $\bf B$ are derived from the improved field strength tensor
\[
F_{\mu\nu}~=~F^{(cl)}_{\mu\nu} - {1 \over 6}(a^2_\mu D^2_\mu~+~a^2_\nu D^2_\nu)~F^{(cl)}_{\mu\nu}~,
\]
where $~F^{(cl)}_{\mu\nu}$ is the standard clover definition. We set $c_i = 1,~i=0,\ldots,8$~.

In this preliminary study we calculate the low-lying $S$-wave triplet and singlet states
$P$-wave singlet states.

For $S$-wave states we use the propagators
\bea
\lefteqn{\mbox{singlet:}~~~~\Gamma^S_a(\bp,t;\by)~=}\nn\\
\lefteqn{~\sum_\bx \cos(\bp\cdot\bx)\mbox{ReTr}\bb G^{S\dagger}_a(\bx,t;\by) G^S_1(\bx,t;\by)\eb,}\nn\\
\lefteqn{\mbox{triplet:}~~~~\Gamma^S_a(\bp,t;\by)~=}\nn\\
\lefteqn{~\sum_\bx \cos(\bp\cdot\bx)\mbox{ReTr}\bb\bsig G^{S\dagger}_a(\bx,t;\by) 
\bsig G^S_1(\bx,t;\by)\eb,}\nn
\eea
where $G^S$ is evolved with a source defined by eqn. \ref{smear_S}, where $\bp$ 
is an appropriate lattice momentum and where the subscripts 
label different initial sources. In practice, we considered only $a=1$ corresponding to the local 
delta-function source and $a=2$ corresponding to smearing by $l=a_s$. 

For $P$-wave states we use the propagator
\bea
\lefteqn{\Gamma^P_a(\bp,t;\by)=}\nn\\
&&\sum_\bx \mbox{ReTr}\bb G_{i,a}^{P\dagger}(\bx,t;\by) \cdot D_iG^S_a(\bx,t;\by)\eb,\nn
\eea
where $G^P$ has been evolved from a source defined by eqn. \ref{smear_P} with label $i$
denoting the spatial derivative. Again, we choose $a=1,2$ defined as for the $S$-wave states.
The averaged propagators are fitted by a full two- or three-exponential fit using the
standard SVD technique with full correlation matrix to give the various mass differences. 
The value of $a_t$ is determined from the $1^1P_1-1^3S_1$ mass difference using the experimental 
value of $440$MeV and then $a_s = a_t/\chi_R$. 

\begin{table*}[hbt]
\setlength{\tabcolsep}{1.5pc}
\newlength{\digitwidth} \settowidth{\digitwidth}{\rm 0}
\catcode`?=\active \def?{\kern\digitwidth}
\caption{\null}
\label{t1}
\begin{tabular*}{\textwidth}{@{}l@{\extracolsep{\fill}}ccccccc}\hline
$L~~$&$m_ba_t$&$(1^1P_1 - 1^3S_1)a_t$&$a_t^{-1}\,$MeV&$a_s^{-1}\,$MeV&
$M_k\chi^2a_t$&$M_k\,$MeV\\\hline
6&1.5&0.1453(8)&3028(16)&794(5)&50.05(16)&10413(100)\\
8&1.5&0.1467(40)&2999(80)&786(20)&48.4(1)&9973(300)\\
8&1.7&0.1584(40)&2780(70)&723(20)&58.5(2)&11174(250)\\
8&1.9&--&--&--&68.7(2)&13122(250)\\\hline
\end{tabular*}
\end{table*}

\begin{table*}[hbt]
\setlength{\tabcolsep}{1.5pc}
\catcode`?=\active \def?{\kern\digitwidth}
\caption{\null}
\label{t2}
\begin{tabular*}{\textwidth}{@{}l@{\extracolsep{\fill}}cccccc}\hline
&L&$\D E\,a_t$&$\D E\,$MeV&$\D E$(isotropic) MeV&$\D E$(expt.) MeV\\\hline
&6&0.00598(2)&18.11(7)&&\\
\rb{$1^3S_1-1^1S_0$~~}&8&0.0060(3)&18.0(1.0)&\rb{26.1(1)}&\rb{--}\\[1.5mm]
&6&0.190(4)&575(12)&&\\
\rb{$2^1S_0-1^1S_0$}&8&0.185(4)&554(20)&\rb{610(50)}&\rb{--}\\[1.5mm]
&6&0.185(3)&560(10)&&\\
\rb{$2^3S_1-1^3S_1$}&8&0.186(4)&{558(20)}&\rb{560(50)}&\rb{560}\\[1.5mm]
&6&0.1453(8)&&&\\
\rb{$1^1P_1-1^3S_1$}&8&0.1467(40)&\rb{440}&\rb{440}&\rb{440}\\[1.5mm]
&6&0.37(1)&1120(33)&&\\
\rb{$2^1P_1-1^3S_1$}&8&0.38(1)&1140(33)&\rb{--}&\rb{775}\\\hline
\end{tabular*}
\end{table*}

In the case of the $1^3S_1$ state the kinetic mass, $M_k$, is determined by fitting the exponent 
of the leading exponential term to
\[
E_S(\bp)~=~E_S(0)~+~p^2 / 2M_k~+~(p^2)^2 / 8M_k^3~.
\]
The natural unit in the simulation for $M_k$ is $a_t/a_s^2 = 1/(a_t\chi_R^2)$. The bare quark mass
entering the definition of $H$, eqn. \ref{H}, is chosen so that this kinetic mass agrees with
the experimental $\Upsilon$ mass. In terms of the computed dimensionless kinetic mass, $\bar{M}_k$,
we have $M_k = \bar{M}_k / a_t\chi_R^2~$. This equation yields a method for measuring the renormalized 
anisotropy $\chi_R$. Given the dimensionless kinetic masses for the these states, $\bar{M}^S_k$ and 
$\bar{M}^P_k$ respectively, we find
\[
\chi_R^2~=~(\bar{M}^P_k~-~\bar{M}^S_k) / (\bar{E}_P~-~\bar{E}_S)~,
\]
where $\bar{E}_P$ and $\bar{E}_S$ are the dimensionless exponents of the leading decay in each 
case. 

The gauge field configurations were generated by using Cabibbo-Marinari and microcanonical
updates.  The simulation was carried out on the Hitachi SR2201 parallel computer at the 
University of Cambridge High Performance Computing Facility. An 8 hour job produced
about 2500 independent configurations and propagator.

\vskip -5truemm


\section{\bf Results and Conclusion}

In Table \ref{t1} we show the results on a $L^3\times T$ lattice for the $1^3S_1$ kinetic
mass as a function of $m_b$ (eqn. \ref{H}) and the determination of the lattice spacing from
$1^1P_1-1^3S_1$. We have not computed the spectrum for $m_b = 1.9$ but the lattice spacings
should change only little from the values quoted for the other masses. We worked
with $m_b = 1.5a_t^{-1}$ for which $M_k$ is a bit high but tolerable.
Our value for $a_s^{-1} \sim 800$MeV agrees well with the value communicated
to us by Alford et al. The $S$ and $P$ state mass differences in the various cases are given
in Table \ref{t2}. In the second to last column are the results from our high statistics analysis of
UKQCD quenched isotropic lattice configurations on $16^3\times 48$ at $\gb = 6.0$ \cite{drea}.
A combined fit to the $1^3S_1$ and $1^1S_0$ data gives the triplet-singlet mass difference.
The fits to both $S$ and $P$ wave data are very good. 

Our preliminary study of NRQCD using an improved action on an anisotropic lattice
has shown that the results for heavy quark states agree well with similar studies using a 
standard action on isotropic lattices of smaller spatial lattice spacing, and the
accuracy with which parameters can be measured with modest amounts of computer time
is superior. There is no lattice size dependence in changing from $L=8$ to $L=6$.

\vskip -4mm
\bibliography{nrqcd_refs}
\bibliographystyle{unsrt}

\end{document}